\documentstyle[psfig]{mn}

\newcommand\pp     {$\pm$}
\newcommand\pers     {s$^{-1}$}
\newcommand\micros  {$\mu$s}

\title[NBO in Cygnus X-2]{Normal-Branch Quasi-Periodic Oscillations
during the High Intensity State of Cygnus X-2}

\author[Rudy Wijnands \& Michiel van der Klis]
   {Rudy Wijnands$^1$\thanks{Chandra Fellow; Email:
   rudy@space.mit.edu} \& Michiel van der Klis$^2$\\
   $^1$Center for Space Research, MIT, 77 Massachusetts Avenue,
       Cambridge, MA 02139-4307, USA\\
   $^2$Astronomical Institute ``Anton Pannekoek'',
       University of Amsterdam,
       Kruislaan 403, NL-1098 SJ Amsterdam, The Netherlands}

\begin{document}
\maketitle

\begin{abstract}
Using data obtained with the {\it Rossi X-ray Timing Explorer}, we
report the detection of a 5 Hz quasi-periodic oscillation (QPO) in the
bright low-mass X-ray binary and Z source Cygnus X-2 during high
overall intensities (the high intensity state). This QPO was detected
on the so-called normal branch and can be identified with the normal
branch QPO or NBO.  Our detection of the NBO is the first one during
times when Cygnus X-2 was in the high intensity state.  The rms
amplitude of this QPO decreased from 2.8\% between 2--3.1 keV to
$<$1.9\% between 5.0--6.5 keV. Above 6.5 keV, its amplitude rapidly
increased to $\sim$12\% rms above 16 keV. The time lags of the QPO
were consistent with being zero below 5 keV (compared to the 2--3.1
keV band), but they rapidly increased to $\sim$70 ms (140\degr) around
10 keV, above which the time lags remained approximately constant near
70 ms.  The photon energy dependencies of the rms amplitude and the
time lags are very similar to those observed for the NBO with other
satellites ({\it Ginga}, {\it EXOSAT}) at different (i.e. lower)
intensity states.

\end{abstract}


\section{Introduction \label{intro}}

\begin{figure}
\begin{center}
\begin{tabular}{c}
\psfig{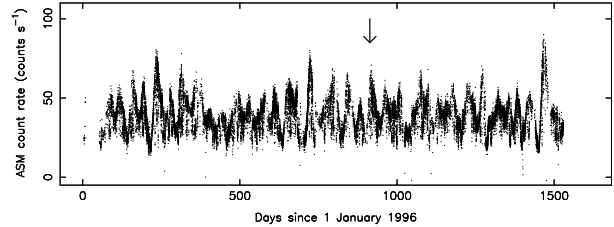}
\end{tabular}
\caption{The {\it RXTE} All Sky Monitor light curve (1.3--12.1 keV)
of Cygnus X-2 clearly showing the long-term X-ray variations. The
arrow indicates when the observations of Cygnus X-2 were taken. The
errors on the count rates are typical 3\%--7\%.
\label{fig:asm_lightcurve} }
\end{center}
\end{figure}

Although the low-mass X-ray binary (LMXB) and Z source (Hasinger \&
van der Klis 1989) Cygnus X-2 is one of the best studied LMXBs, its
complex behaviour is far from understood. On time-scales of hours to
days, the source traces out its characteristic Z shape pattern in the
X-ray colour-colour diagram (CD). The branches of this Z are called,
from top to bottom, the horizontal branch (HB), the normal branch
(NB), and the flaring branch (FB). On time-scales of weeks to months,
the overall intensity varies smoothly by a factor of about 4 (see,
e.g., Fig.~\ref{fig:asm_lightcurve}). The exact morphology of the Z
track and the position of this track in the CD (and hence the overall
hardness of the X-ray spectrum) varies significantly when the overall
intensity changes (see, e.g., Kuulkers, van der Klis, \& Vaughan 1996
and Wijnands et al. 1997 for overviews of these changes).  The first
few months of the {\it Rossi X-ray Timing Explorer} ({\it RXTE};
Bradt, Rothschild, \& Swank 1993) All Sky Monitor (ASM) data of Cygnus
X-2 suggested an $\sim$78 day period in the long-term variations
(Wijnands, Kuulkers, \& Smale 1996). Although this period is still one
of the strongest periods, analyses of all {\it RXTE}/ASM data now
available (up to 2000 October 9) show that the long-term variations
are more complex than only one single periodicity (see also Kong,
Charles, \& Kuulkers 1998 and Paul, Kitamoto, \& Makino 2000).

\begin{table*}
\caption{Log of the {\it RXTE} observations of Cygnus X-2\label{tab:log}}
\begin{flushleft}
\begin{tabular}{llll}
\hline
Observation ID & Start & End  & On source time\\
               & (UT)  & (UT) & (ks)\\
\hline
30418-01-01-00 & 1998 July 2 10:48 & 1998 July 2 14:30 & 8.2 \\  
30418-01-02-00 & 1998 July 3 11:37 & 1998 July 3 12:42 & 3.3 \\ 
30418-01-02-01 & 1998 July 3 13:13 & 1998 July 3 14:15 & 3.7 \\ 
30418-01-03-00 & 1998 July 4 11:38 & 1998 July 4 14:15 & 7.3 \\ 
30418-01-04-00 & 1998 July 5 11:38 & 1998 July 5 14:14 & 7.3 \\ 
30418-01-05-00 & 1998 July 6 09:07 & 1998 July 6 14:14 & 11.7 \\ 
\hline
\end{tabular}
\end{flushleft}
\end{table*}

The physical mechanism behind the long-term variations in Cyg X-2 is
unknown. Several mechanisms have been proposed, such as variations in
the mass accretion rate onto the neutron star, a precessing accretion
disc, or a precessing neutron star (see Wijnands et al. 1997 and
references therein for a discussion of these possible
mechanisms). Although variations in the mass accretion rate are
unlikely to cause the long-term variability (these variations are
thought to produce the motion of Cyg X-2 along the Z track; see, e.g.,
Hasinger \& van der Klis 1989, Kuulkers et al. 1996, and Wijnands et
al. 1997), a precessing accretion disc or a precessing neutron star
are still possible. Besides affecting the X-ray count rate and the
X-ray spectrum, both mechanisms would most likely also affect the
rapid X-ray variability, although it is unclear how exactly the
variability would be affected. A detailed study of the rapid X-ray
variability during different overall intensities might give new
insights into the physics behind the long-term variability of Cyg
X-2. However, such a detailed study has not yet been performed. Below
we briefly list the results so far available.

On the HB, quasi-periodic oscillations (QPOs) between 15 and 55 Hz are
observed (the horizontal branch QPOs or HBOs).  These QPOs are also
often observed on the upper part of the NB. A detailed comparison of
the HBO properties during different overall intensities (hereafter
referred to as different intensity states) has not yet been
made. Although Wijnands et al. (1997) reported that the HBO is always
observed on the HB regardless of the overall intensity of Cygnus X-2,
the HBO properties could differ in detail between different intensity
states.

On the NB, QPOs with frequencies near 5 Hz can also be observed (the
normal-branch oscillations or NBOs; sometimes observed simultaneous
with the HBO). Also the NBO properties have not yet been compared
between different intensity states, although Wijnands et al. (1997)
did not detect the NBO (on the lower NB; with typical upper limits of
$<$1\% rms on the NBO amplitude) during the high intensity states
observed using {\it Ginga} data of Cyg X-2. The NBOs {\it were}
observed during lower intensity states.

Another type of QPO (with frequencies between 300 and 900 Hz) was
recently discovered (Wijnands et al. 1998) in Cygnus X-2 when this
source was on the HB during a medium intensity state: the kHz
QPOs. These QPOs were not observable during a higher intensity state
(Smale 1998) with upper limits which were smaller than the measured
values.

Here, we report the first detection of a NBO in Cyg X-2 during a high
intensity state demonstrating that this type of QPO is so far always
observable regardless of the overall intensity of the source.  The
energy dependence of the rms amplitude and the time lags of this NBO
are very similar to those observed during lower intensity states.

\section{Observations and  Analysis  \label{observations}}

Cygnus X-2 was observed by the {\it RXTE} proportional counter array
(PCA; Jahoda et al. 1996) on several occasions between 1998 June 2--7
for a total of $\sim$42 ksec (see Table~\ref{tab:log} for a log of the
observations).  Data were obtained in 129 photon energy channels
(covering 2--60 keV) with a time resolution of 16 s. Simultaneous data
were obtained in two ``Single Bit'' modes (one energy channel each;
time resolution of 128 \micros; energy range 2--5.0 and 5.0--13.0
keV), one ``Event'' mode (64 channels; 16 \micros; 13.0--60 keV), and
one ``Binned'' mode (16 channels; 2 ms; 2--13.0 keV). We used the 16 s
data to create the CD and the hardness-intensity diagram (HID; see the
caption of Fig.~\ref{fig:cd_hid} for the energy bands used to
calculate the colours). We used the Single Bit modes and the Event
modes to create 1/16--2048 Hz power spectra to study the rapid X-ray
variability above 100 Hz (i.e., to search for kHz QPOs); we used the
Binned and the Event modes to create 1/16--256 Hz power spectra and
cross spectra to study the rapid X-ray variability below 100 Hz (i.e.,
to study the NBO and HBO). The strengths of the power spectral
components are all given for the energy range 2--60 keV, unless
otherwise noted.

\section{Results \label{results}}

\begin{figure}
\begin{center}
\begin{tabular}{c}
\psfig{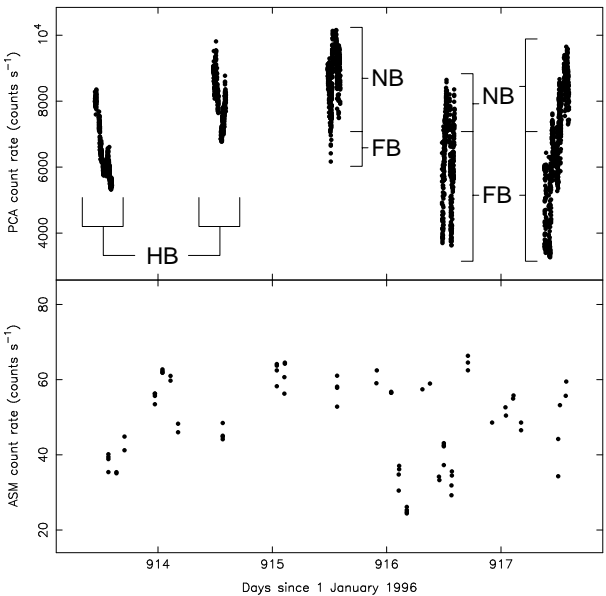}
\end{tabular}
\caption{The {\it RXTE}/PCA light curve (2.0--16.0 keV) ({\it top
panel} ) and the All Sky Monitor light curve (1.3--12.1 keV) ({\it
bottom panel}) at the time of the observations of Cygnus X-2. In the
{\it top panel}, it is indicated on which branch the source was during
the {\it RXTE}/PCA observations.  The errors on the PCA and the ASM
count rates are typical 0.3\%--0.5\% and 3\%--7\%, respectively.
\label{fig:asmpca_lightcurve} }
\end{center}
\end{figure}

Figure~\ref{fig:asm_lightcurve} shows the {\it RXTE}/ASM light curve
of Cyg X-2 clearly showing the long-term X-ray variations.  During the
observations presented here, Cygnus X-2 was at high count rates.  The
{\it RXTE}/ASM count rates were approximately 60 counts
\pers~(Fig.~\ref{fig:asmpca_lightcurve}). The upper envelope of the
ASM light curves in Figures~\ref{fig:asm_lightcurve} and
\ref{fig:asmpca_lightcurve} corresponds to epochs during which Cyg X-2
was on the NB; the short-lasting ``drop-outs'' from this upper
envelope are excursions into the HB and the FB. These excursions do
not define a lower envelope demonstrating that the overall intensity
of Cyg X-2 is mainly defined by the count rate on the NB.

\begin{figure}
\begin{center}
\begin{tabular}{c}
\psfig{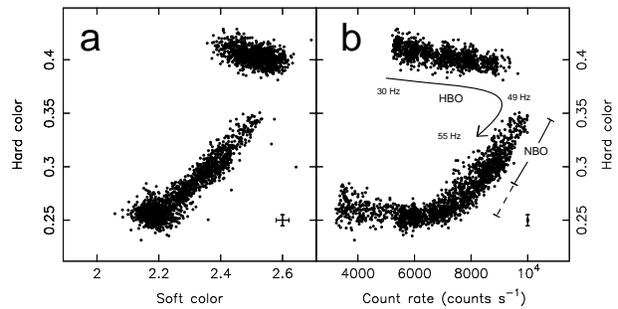}
\end{tabular}
\caption{Colour-colour diagram ({\it a}) and hardness-intensity
diagram ({\it b}) of Cygnus X-2.  The soft colour is the count rate
ratio between 3.5--6.4 and 2.0--3.5 keV, the hard colour that between
9.7--16.0 and 6.4--9.7 keV, and the count rate is in the energy range
2.0--16.0 keV. All points are 16 s averages. In ({\it b}) it is
indicated when the HBO and the NBO were present in the data and how
the frequency of the HBO changed (indicated by the arrow). The solid
line labelled NBO indicates where the FWHM of the NBO was smaller than
5 Hz. All the data corresponding to this part of the Z track were used
to calculate the energy dependence of the rms amplitude and the time
lags of the NBO.  The dashed line indicates where the FWHM increased
to 7--13 Hz. Typical error bars on the colours and the intensity are
indicated in the lower right corners of the diagrams.
\label{fig:cd_hid}}
\end{center}
\end{figure}

The CD and HID of the data are shown in Figure~\ref{fig:cd_hid}. In
both diagrams, clear horizontal and normal branches are visible. The
FB is not clearly visible in the CD (only as a broadening of the lower
part of the normal branch) but in the HID an extended FB is
present. The count rate decreased when Cygnus X-2 entered the FB.  We
selected the power spectra based on the position of the source on the
track in the HID. The HID was used instead of the CD because by using
the HID it was possible to disentangle the power spectra corresponding
to the lower part of the NB and those corresponding to the FB.

\begin{figure}
\begin{center}
\begin{tabular}{c}
\psfig{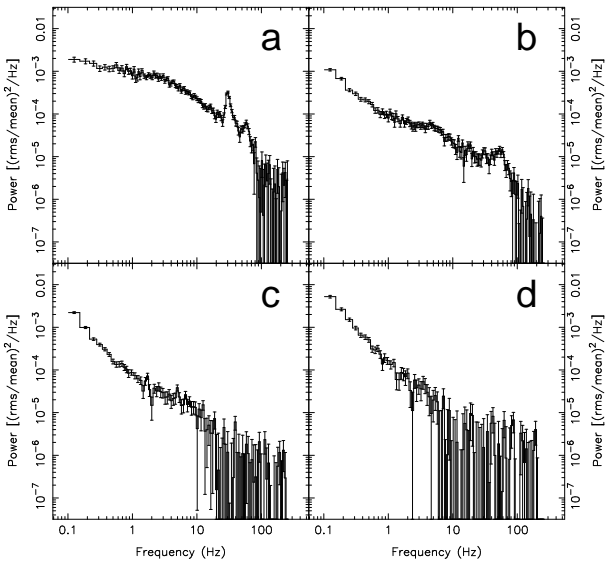}
\end{tabular}
\caption{Typical power spectra for the full 2--60 keV energy range
of Cygnus X-2 on: ({\it a}) the horizontal branch, ({\it b}) the upper
part of the observed normal branch, ({\it c}) the normal-flaring
branch vertex, and ({\it d}) the end of the flaring branch. The
Poisson level has been subtracted.
\label{fig:powerspectra_z}}
\end{center}
\end{figure}

In Figure~\ref{fig:powerspectra_z}, typical 2--60 keV power
spectra are shown for different positions of the source on the Z
track. On the HB, clear HBOs, their second harmonics, and strong
(8\%--10\% rms) band-limited noise (the Z source low-frequency noise
or LFN; Fig.~\ref{fig:powerspectra_z}{\it a}) were present. The
frequency of the HBO was $\sim$30 Hz at the left end of the HB and
increased smoothly to 49 Hz at the right end (see also
Fig.~\ref{fig:cd_hid}{\it b}). The rms amplitude of the HBO at the
left end of the HB was 5\% and decreased to 4\% when the source moved
slightly to the right on the HB. The amplitude of the HBO on the rest
of the HB stayed approximately constant at 4\% rms. The FWHM of the
HBO was 5 Hz at the left end of the HB and increased to 20 Hz when the
source moved to the right end. Together with the HBO, its second
harmonic could be detected from the left end up to halfway the HB,
with an rms amplitude decreasing from 4.4\% to 2.6\% and a FWHM
increasing from 24 to 36 Hz.

On the upper part of the observed normal branch
(Fig.~\ref{fig:powerspectra_z}{\it b}), the HBO was still visible
around 55 Hz (with an rms amplitude of 2\%--3\% and a FWHM of 20--30
Hz) together with a noise component below 1 Hz following a power law
(the Z source very-low frequency noise or VLFN) and a peaked noise
component near 5 Hz. Further down the NB and at the beginning of the
FB (Fig.~\ref{fig:powerspectra_z}{\it c}), no QPOs were present, but
the VLFN still was, together with a weak (several percent) noise
component following a cutoff power law (cutoff frequency near 10 Hz;
probably the Z source high-frequency noise or HFN). On the left end of
the FB only the VLFN remained (Fig.~\ref{fig:powerspectra_z}{\it d}).

\begin{figure}
\begin{center}
\begin{tabular}{c}
\psfig{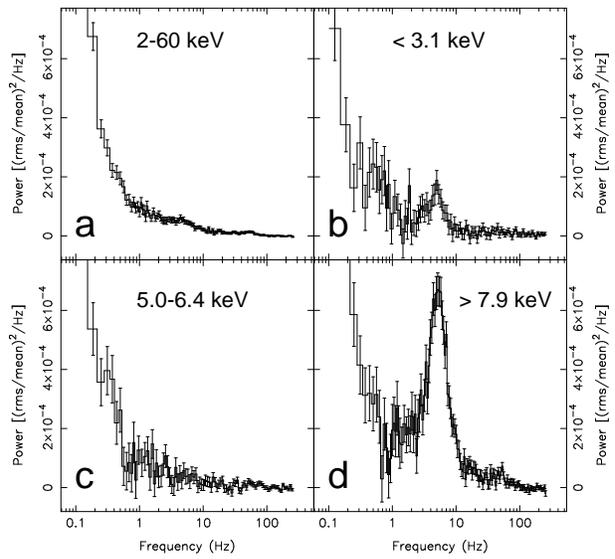}
\end{tabular}
\caption{Typical power spectra of Cygnus X-2, when the source was
on the upper part of our observed normal branch, in different energy
ranges: ({\it a}) 2--60 keV, ({\it b}) below 3.1 keV, ({\it c})
5.0--6.4 keV, and ({\it d}) above 7.9 keV. The Poisson level has been
subtracted.  The 2--60 keV power spectrum ({\it a}) is the same as
shown in Fig.~\ref{fig:powerspectra_z}{\it b}, but now with the power
axis plotted linear.
\label{fig:powerspectra_nbo}}
\end{center}
\end{figure}

So, when using power spectra created for the total {\it RXTE}/PCA
photon energy range (2--60 keV), no NBO were detected on the
NB. However, when we examined the peaked noise component on the upper
part of the observed NB (Fig.~\ref{fig:powerspectra_z}{\it b}) in more
detail and in different energy ranges, a clear QPO around 5 Hz
emerged. Figure~\ref{fig:powerspectra_nbo} shows the power spectra
corresponding to the upper normal branch but at different
energies. The power spectrum in Figure~\ref{fig:powerspectra_nbo}{\it
a} is the same as the one presented in
Figure~\ref{fig:powerspectra_z}{\it b}, although now plotted with the
power axis in linear scale in order to show the QPO more
clearly. Figures~\ref{fig:powerspectra_nbo}{\it b} to {\it d} show the
corresponding power spectra at different energies. At energies above
7.9 keV a very significant (29 $\sigma$) QPO is clearly visible with a
frequency of 5.37$^{+0.08}_{-0.03}$ Hz, a FWHM of 4.4\pp0.2 Hz, and an
rms amplitude (7.9--60 keV) of 6.7\%\pp0.1\%. Note that the FWHM of
the QPO is such that according to the usual criterion for a noise
component to be called a QPO (the ratio of the FWHM to the frequency
is $<$ 0.5), this NBO was not truly a QPO. However, below we will use
the term QPO for this peaked noise feature.  By comparing
Figures~\ref{fig:powerspectra_nbo}{\it a} and {\it d}, it is clear
that this QPO is only visible as a broad peaked noise component near 5
Hz when combining the data of the whole {\it RXTE}/PCA energy
range. The QPO was also visible at low ($<$5 keV) photon energies
(Fig.~\ref{fig:powerspectra_nbo}{\it b}), but not at energies between
5.0 and 6.4 keV.

After discovering the NBO at high photon energies, we reexamined all
the power spectra corresponding to the different locations of the
source on the Z track. We used the 7.9--60 keV energy range instead of
the 2--60 keV one. The 7.9--60 keV power spectra on the horizontal
branch were very similar to those in the 2--60 keV band. As said
above, on the upper part of the observed normal branch a broad NBO was
now clearly visible. The frequency (5.4 Hz), the FWHM (4 Hz), and the
rms amplitude (7\%) of the NBO did not change significantly when the
source moved down the NB. Halfway down the NB, the QPO became much
broader (with a FWHM of 7--13 Hz) and the NBO decreased slightly in
amplitude (to 5\%--6\% rms).  On the lower part of the NB (near the
NB-FB vertex), the NBO could not be detected anymore with upper limits
on the rms amplitude of 3\%--4\% (see also Fig.~\ref{fig:cd_hid}{\it
b} for the region where the NBO could be detected). At the NB-FB
vertex and on the FB the 7.9--60 keV power spectra were very similar
to the 2--60 keV power spectra.

\begin{figure}
\begin{center}
\begin{tabular}{c}
\psfig{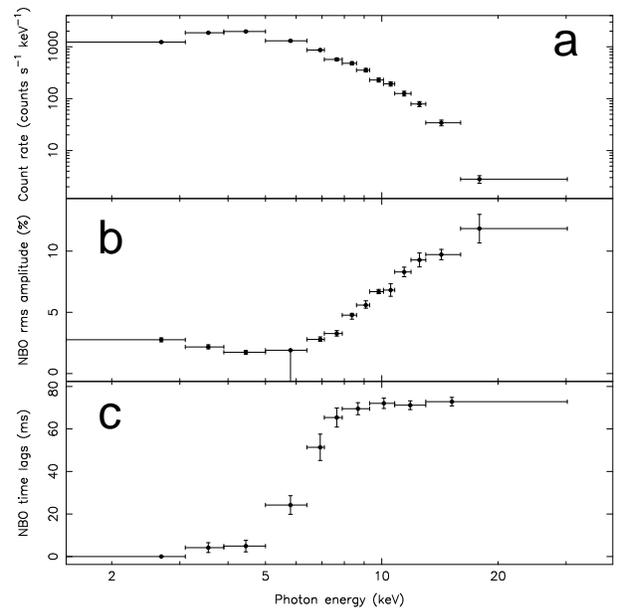}
\end{tabular}
\caption{The count rate of the upper part of the observed NB ({\it
a}), the rms amplitude of the NBO ({\it b}), and the time lags of the
NBO ({\it c}) as a function of photon energy. The errors on the photon
energy range from the averaged mean photon energy in the energy
channels to the boundary of these channels.
\label{fig:timelags}}
\end{center}
\end{figure}

Figure~\ref{fig:timelags}{\it a} shows the count rate in the
energy bands used to calculate the rms amplitude of the NBO.  The rms
amplitude of the NBO as a function of photon energy is plotted in
Figure~\ref{fig:timelags}{\it b}. In order to calculate these rms
amplitudes, we only used the data of the upper part of the observed NB
for which the NBO had a FWHM which was lower then 5 Hz (see also
Fig.~\ref{fig:cd_hid}{\it b}).  The NBO decreased from about 3\% rms
at the lowest energies to $<$2\% rms between 5 and 6.4 keV. Above 6.4
keV, the strength of the NBO rapidly increased to 10\% at 15 keV. At
higher energies the strength seemed to level of, although a firm
conclusion cannot be made.

For the same data, we determined the time lags of the NBO between the
different energy bands. In order to calculate the time lags of the
NBO, we calculated the lags in the 4.4 Hz interval (corresponding to
the FWHM of the NBO) centred on the peak frequency of the QPO (5.37
Hz). We subtracted the average cross-vector between 80 and 250 Hz in
order to minimize the dead-time effects on the lags induced by Poisson
fluctuations (see van der Klis et al. 1987). The resulted time lags
are presented in Figure~\ref{fig:timelags}{\it c}. We used the 2--3.1
keV energy band as reference band. The time lags between this
reference band and the bands below 5 keV were consistent with
zero. Above 5 keV, the time lags increase rapidly to $\sim$70 ms for
energies between 8 and 9 keV. At higher energies, the time lags
remained constant near 70 ms. This means that the hard photons lag the
soft photons by about 70 ms (2.4 radians or 140\degr).

We searched for kHz QPOs in the total energy band and in the energy
range 5.0--60 keV (as used by Wijnands et al. 1998 when kHz QPOs were
present in Cyg X-2). None were found with typical (assuming a FWHM of
150 Hz) 95\% confidence upper limits of 2\%--3\% (1\%--2\%) on the HB,
2\%--3\% (1\%--1.5\%) on the NB, and 2.5\%--3.5\% (1.5\%--2.0\%) on
the FB in the energy range 5.0--60 keV (the limits enclosed within
parentheses are for the total {\it RXTE}/PCA energy range). Two X-ray
bursts were observed. We searched for (nearly) coherent oscillations,
as reported in several other LMXBs (e.g., Strohmayer et al. 1996;
Smith, Morgan, \& Bradt 1997), but none were found with upper limits
of 4\%--7\% (2--60 keV).

\section{Discussion \label{discussion}}

We have detected the NBO on the upper part of our observed normal
branch of Cygnus X-2 when this source was at relative high overall
intensities (Fig.~\ref{fig:asm_lightcurve}). The shape of the Z track
in the CD and the HID was very similar to those observed with {\it
Ginga} for Cygnus X-2 in the high intensity state (Wijnands et
al. 1997; i.e., compare their Fig.~1 with our
Fig.~\ref{fig:cd_hid}). Therefore, we conclude that Cygnus X-2 was in
a high overall intensity state during the observations reported
here. This conclusion is strengthened by the fact that our power
spectra on the lower normal branch resemble those obtained by Wijnands
et al. (1997) in the high intensity state, especially the
non-detection of the NBO on the lower part of the NB (compare their
Figs.~8{\it g--i} with our Figs.~\ref{fig:powerspectra_z}{\it c} and
{\it d}).  Wijnands et al. (1997) were unable to search for the NBO
further upwards on the NB (the middle and upper part of the NB) due to
the fact that they did not have enough high time resolution data at
those positions on the Z track. Our {\it RXTE}/PCA data are the first
high time resolution data available on the middle and upper part of
the NB of Cygnus X-2 during the high intensity state.

Our first clear detection of the NBO in the high intensity state of
Cygnus X-2 shows that all power spectral components which are observed
below 100 Hz in Cygnus X-2 are present at high overall intensities.
So far, the only clear difference in the timing properties of Cygnus
X-2 during different intensity states are the detection of the kHz
QPOs on the HB during a medium intensity state (Wijnands et al. 1997)
but not during high intensity states (Smale 1998 and this paper). More
detections of the kHz QPOs in Cygnus X-2 are needed in order to
determine the exact reason(s) why the kHz QPOs are only present during
preferred intensity states.

Wijnands et al. (1997) pointed out that a precessing accretion disc,
one of the proposed explanations for the long-term X-ray variations of
Cyg X-2 (e.g., Kuulkers et al. 1996; Wijnands et al. 1996), could not
easily account for the non-detection of the NBO during the high
intensity state. Our detection of the NBO during this state now shows
that the arguments used by Wijnands et al. (1997) are not valid any
more when only considering the NBO. However, when also considering the
kHz QPOs, it is clear that the precessing accretion disc model has
still difficulties explaining the difference in rapid X-ray timing
variability (i.e., the kHz QPOs) when Cyg X-2 is at different
intensity states.  In the precessing accretion disc scenario, it is
possible that more matter is blocking the emission region during low
intensity states (hence the lower overall intensity) than during high
intensity states. If this the case then during these low states, the
amplitudes of the kHz QPOs should be significantly lower than in the
high states due to more scattering of the radiation. However, we see
exactly the opposite: no kHz QPOs are detected during the high
intensity state. It is also possible that the precession of the
accretion disc causes changes in the projected area of the inner disc
which could result in the different intensity states. However, then
the kHz QPOs would be expected in all intensity states, i.e., also
during the high intensity state.  The non-detections of the kHz QPOs
during the high intensity state constitute a serious problem for the
precessing accretion disc model.

We now compare our results obtained for the NBO to those which have
been published previously. The best cases to compare our results with
are the {\it Ginga} June 1987 observations of Cygnus X-2 reported by
Mitsuda \& Dotani (1989; see also Wijnands et al. 1997) and the {\it
EXOSAT} November 14/15 1985 observations which are reported by Dieters
et al. (2000; see also Hasinger et al. 1985,1987). Wijnands et
al. (1997) classified the intensity state of Cyg X-2 during the {\it
Ginga} observations as intermediate between the high and the medium
intensity state. Kuulkers et al (1996) classified the state during the
{\it EXOSAT} observations as the medium state (see Wijnands et
al. 1997 for the classification of the different states of Cygnus
X-2).

During the {\it Ginga} observations (Mitsuda \& Dotani 1989; Wijnands
et al. 1997) and the {\it EXOSAT} observations (Dieters et al. 2000),
the NBO was best visible on the middle part of the NB. The NBO was
much less clear on the upper and lower parts of the NB (although
excess power on top of a power-law function was present near 5 Hz).
The exact position on the Z track where the NBO is present during
those observations is very similar to the positions on the track where
we detect the NBO.  Although we detected the NBO up to the upper most
part of our observed NB and Mitsuda \& Dotani (1989) and Dieters et
al. (2000) did not detect the NBO on the upper part of their observed
NB, this difference is most likely due to the fact that we did not
observe the NB all the way to the HB/NB vertex (see
Fig.~\ref{fig:cd_hid}{\it b}) and most likely they did.  If we would
have observed the NB all the way to the HB/NB vertex most likely we
would not have detected the NBO either on this part of the Z
track. Wijnands et al. (1997) also reported that the NBO was best
visible at the middle of the NB during the medium state and that it
tends to get broader when the source moves further down the NB. The
energy dependence and time lags of our NBO are also nearly identical
to those obtained for the NBO by Mitsuda \& Dotani (1989) and Dieters
et al. (2000). This indicates that the physical mechanism behind
the long-term variation of the X-ray flux of Cygnus X-2 does not
significantly affects the properties of the NBO.

From the presently available results, we conclude that so far no
significant difference is present in the NBO properties during the
high intensity state of Cygnus X-2 and during its intermediate and
medium intensity state.  Recently, Kuulkers, Wijnands, \& van der Klis
(1999) reported on {\it RXTE} observations of Cyg X-2 when this source
was in a very low intensity state. No NBOs were detected. However, no
complete Z track was traced out during these observations but most
likely only the lower part of the NB and part of the FB. It is
possible that when a complete Z track would have been traced out, the
NBO would have been present at the middle part or upper part of the
NB.

To explain the NBO in Z sources, it has been proposed that, when those
sources are accreting close to the Eddington accretion limit, part of
the accretion will occur in an approximately spherical radial inflow.
In this radial flow a radiation pressure feed-back loop can be set up
that causes 5--7 Hz oscillations in the optical depth of the flow
(Fortner et al. 1989; Lamb 1989; Miller \& Lamb 1992). This
radiation-hydrodynamic model is able to explain the combination of the
minimum in the rms amplitude spectrum of the NBO and the
$\sim$140--150\degr~phase shift in the phase lags at the same photon
energy (Miller \& Lamb 1992). Miller \& Lamb (1992) used the results
reported by Mitsuda \& Dotani (1989) to constrain their model. Because
our results are so similar to those published by Mitsuda \& Dotani
(1989), we are unable to constrain this radiation-hydrodynamic model
further. In order to constrain this model, data are needed with a
higher energy resolution than our {\it RXTE}/PCA data and/or data have
to be obtained for the energy range below 3 keV, a region which we are
unable to probe with our {\it RXTE}/PCA data.  However, it has already
been showed that this model cannot easily explain the differences in
the NBO properties observed for Cyg X-2 and those observed for GX 5--1
and Sco X-1 (see, e.g., Dieters et al. 2000).

The recent discovery of a QPO near 7 Hz in the atoll source 4U
1820--30 (Wijnands, van der Klis, \& Rijkhorst 1999) might even impose
a more serious problem to all NBO models.  The properties of this QPO
in 4U 1820--30 (i.e., its frequency and its presence only at the
highest observed mass accretion rates in this source) are similar to
the NBO in the Z sources. However, if indeed these QPOs are the same
phenomenon, then the models explaining the NBO in Z sources which
require near-Eddington mass accretion rate, will not hold because the
highest mass accretion rate observed in 4U 1820--30 (i.e., the
accretion rate at the time of the 7 Hz QPO) is significantly lower
than the critical Eddington mass accretion rate.  This is not only
true for the radiation-hydrodynamic model of Fortner et al. (1989),
but also for alternative models, such as the sound wave model of Alpar
et al. (1992). Clearly, the NBO models have to be significantly
adjusted in order to explain all NBO properties and to explain the
similarities between the NBO and the 7 Hz QPO in 4U 1820--30.

\section*{Acknowledgments}

This work was supported in part by the Netherlands Foundation for
Research in Astronomy (ASTRON) grant 781-76-017, by the Netherlands
Research School for Astronomy (NOVA), and the NWO Spinoza grant 08-0
to E. P. J. van den Heuvel. Support for this work was provided by the
NASA through the Chandra Postdoctoral Fellowship grant number
PF9-10010 awarded by the Chandra X-ray Center, which is operated by
the Smitsonian Astrophysical Observatory for NASA under contract
NAS8-39073.  This research has made use of data obtained through the
High Energy Astrophysics Science Archive Research Center Online
Service, provided by the NASA/Goddard Space Flight Center.

%\clearpage


\begin{thebibliography}{}


\bibitem{}Alpar M.A., Hasinger G., Shaham J., Yancopoulos S., 1992,
A\&A, 257, 627

\bibitem{}Bradt H.V., Rothschild R.E., Swank J.H., 1993, A\&AS, 97, 355

\bibitem{}Dieters S.W., Vaughan B.A., Kuulkers E., Lamb F.K., van der Klis M., 2000, A\&A, 353, 203 

\bibitem{}Fortner B., Lamb F.K., Miller G.S., 1989, Nature, 342, 775

\bibitem{}Hasinger G., Langmeier A., Sztajno M., Pietsch W., Gottwald M., 1985, IAUC 4153

\bibitem{}Hasinger G., in Helfand D.J., Huang J.-H. (eds) 1987
``The Origin and Evolution of Neutron Stars'', IAU Symposium 125, p. 333 

\bibitem{}Hasinger G., \& van der Klis M. 1989, A\&A, 225, 79

\bibitem{}Jahoda K., Swank J.H., Giles A.B., Stark M.J., Strohmayer
T., Zhang W., Morgan E.H.,  1996, SPIE, 2808, 59

\bibitem{}Kong A.K.H., Charles P.A., Kuulkers E., 1998, NewA 3, 301

\bibitem{}Kuulkers E., van der Klis M., Vaughan B.A., 1996, A\&A, 311, 197

\bibitem{}Kuulkers E., Wijnands R., van der Klis M., 1999, MNRAS, 308, 485

\bibitem{}Lamb F.K., 1989, in Proc. 23 rd ESLAB Symp. on ``Two Topics
in X-ray Astronomy'', Bologna, Italy, 13--20 September 1989 (ESA
SP-296), p. 215

\bibitem{}Miller G.S. \& Lamb F.K., 1992, ApJ, 388, 541

\bibitem{}Mitsuda K., Dotani T., 1989, PASJ 41, 531    

\bibitem{}Paul B., Kitamoto S., Makino F., 2000, ApJ 528, 410

\bibitem{}Smale A.P., 1998, ApJ 498, L141

\bibitem{}Smith D.A., Morgan E.H., Bradt H., 1997, ApJ 479, L137

\bibitem{}Strohmayer T.E., Zhang W., Swank J.H., Smale A., Titarchuk
L., Day C., Lee U., 1996, ApJ 469, L9

\bibitem{}van der Klis M., Hasinger G., Stella L., Langmeier A, van
Paradijs J., Lewin W.H.G., 1987, ApJ 319, L13

\bibitem{}Wijnands R.A.D., Kuulkers E., Smale A.P., 1996, ApJ 473, L45

\bibitem{}Wijnands R., van der Klis M., Kuulkers E., Asai K., Hasinger
G., 1997, A\&A, 323, 399

\bibitem{}Wijnands R., et al. 1998, ApJ 493, L87

\bibitem{}Wijnands R., van der Klis M., \& Rijkhorst E.-J., 1999, ApJ,
512, L39


\end{thebibliography}
\end{document}